\documentclass[aps,prl,twocolumn,groupedaddress,showkeys]{revtex4}
\usepackage[pdftex]{graphicx}
\usepackage{amsmath}

\newcommand{\threepartdef}[6]
{
	\left\{
		\begin{array}{lll}
			#1 & \mbox{if } #2 \\
			#3 & \mbox{if } #4 \\
			#5 & \mbox{if } #6
		\end{array}
	\right.
}

\begin{document}


\title{Stability of Superconducting Power-Law Cable-In-Conduit Conductors}


\author{A.Anghel}
\email[]{alexander.anghel@psi.ch}
\affiliation{Paul Scherrer Institute, CH-5232 Villigen-PSI, Switzerland}


\begin{abstract}
The stability properties of cable in conduit conductors with a power-law current-voltage characteristic are investigated using a previously developed model description for the take-off properties of these conductors. The numerical investigation of the transients shows a predictable quench (take-off) behavior of power-law conductors in the frame of a dc phase diagram. Differences between heat pulses of increasing duration, increasing power and different pulse form (square and sinusoidal) are discussed.  
\end{abstract}

\pacs{}

\keywords{superconductor cable, stability power-law volt-ampere characteristic}

\maketitle

\subsection{The quench model and the stability issue}
Recently, a model was developed \cite{anghel1} to explain the take-off behavior of power-law conductors \cite{bruzzone1}. Although this model was restricted to dc or quasi-dc experiments like the critical current or current-sharing temperature measurements where either the the current or the helium temperature are increased slowly in a quasi-stationary manner up to the take-off point, it became soon clear that this analysis is more general such as to include also the transient behavior i.e. the stability issue.

Indeed, the basic equation of the above mentioned dc model

\begin{equation}
	G=H
\end{equation}

which equates the heat generation in the strand, in the power-law formulation (also known as the index heating) 

\begin{eqnarray}\nonumber
	G\equiv G(T_{cond}, T_{He})=E(T_{cond},I_{op})I_{op}=\\
	=E_c\left[\frac{I_{op}}{I_c(T_{cond},B)}\right]^{n}I_{op}
\end{eqnarray}

to the helium cooling

\begin{equation}
H\equiv H(T_{cond},T_{He})=hp_w(T_{cond}-T_{He})
\end{equation}

is nothing else but the steady-state solution of the time-dependent 0D equation

\begin{equation}
\rho SC\frac{\partial T_{cond}}{\partial t}=G-H
\end{equation}

where: $T_{cond}$ and $T_{He}$ are the conductor and helium temperature, $I_{op}$ is the operation current. $E_c$ and $I_c$ the critical electrical field and the critical current (a function of conductor temperature and external magnetic field $B$, and $n$, the power-law index. $\rho$ is the density, $S$ the conductor cross-section area and $C$ the conductor specific heat. The heat exchange coefficient $h$ and the wetted perimeter $p_w$ are used to describe the heat convection at the conductor-helium interface.
Remembering that stability means the ability to return to a stable steady-state after a perturbation has been applied to the system and expressing Eq.1 as

\begin{equation}
T_{He}=T_{cond}-\frac{E_{c}I_{op}}{hp_w}\left[\frac{I_{op}}{I_c(T_{cond},B)}\right]^n
\end{equation}

\begin{figure}
\includegraphics[width=8.3cm]{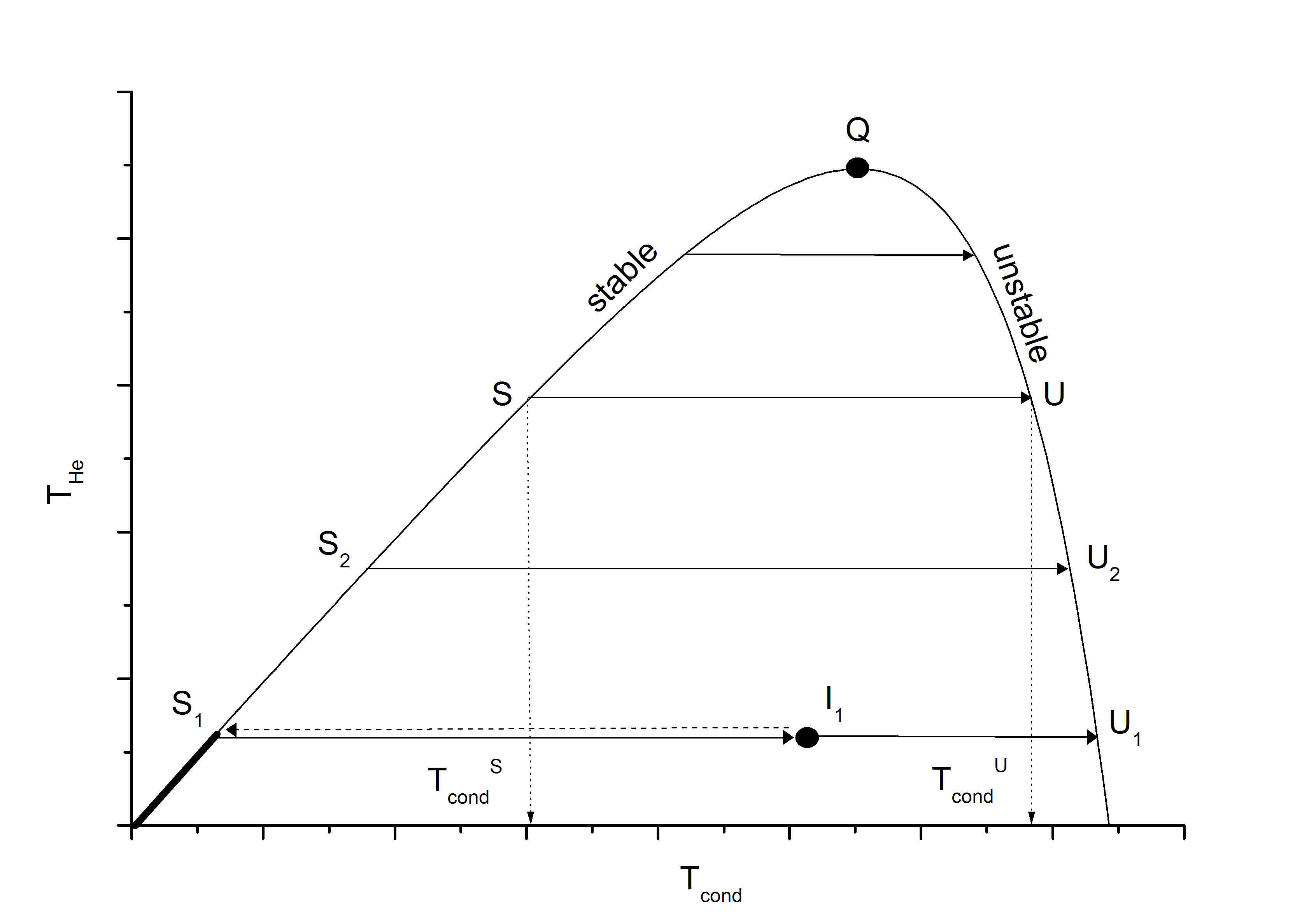}
\caption{Stability phase diagram for a given operating current $I_{op}$ and external field $B$ obtained solving Eq.1 respectively plotting $T_{He}$ as a function of $T_{cond}$ according to Eq.5.}
\label{fig:Bild1}
\end{figure}

we get a kind of phase diagram if we plot $T_{He}$ as a function of $T_{cond}$  for a  given $I_{op}$ as shown in Fig.\ref{fig:Bild1}. It was shown in \cite{anghel1} that the maximum of this function, the point $Q$ in Fig.\ref{fig:Bild1}, defines the quench point and that the left curve consists of pairs of values $(T_{cond},T_{He})$  for which the conductor is in a stable steady-state condition. The right curve is made of unstable points because for every, no matter how small, perturbation, the system do not returns to the initial state. It either diverges, $T_{cond}\rightarrow \infty$ if the initial point is to the right of it or it returns to a point on the left side of the curve, if the initial point was in between the two branches.

Now imagine we have at $t<0$ a conductor in stable condition represented by some point $S_1$ on the left curve of Fig.\ref{fig:Bild1} and apply at $t=0$ a perturbation $P_{pulse}(t)$ of finite duration and magnitude, ignoring for the time being the heat transfer to helium. During the pulse the conductor temperature increases and eventually, at the end of the pulse, it reaches a maximum represented by the point $I_1$ on the diagram in Fig.\ref{fig:Bild1}. It is clear that if the pulse power is such that $I_1$ doesn't reach the right curve or goes beyond it, the system will return to the state $S_1$ after the transient. If the perturbation is such that the conductor temperature during the pulse, goes beyond the right curve, the conductor will quench. The limiting situation is when the pulse power is exactly that necessary to put the conductor temperature on the right curve represented by the point $U_1$ and this defines the stability limit of the conductor. If we had chosen a higher helium temperature, the stable initial condition would be represented by the point $S_2$ and the stability limit would be $U_2$ with the property that the maximum allowed excursion in the conductor temperature $S_2U_2$ is smaller and the limiting pulse power is lower. In the limit, when the helium temperature is such that the initial point is $Q$ the conductor will quench at zero power i.e. $Q$ is indeed the quench point. 

In conclusion, it seems that one does not need to solve the differential equation describing the transient in order to get information on the stability of a conductor. It is enough to draw  the "phase diagram" as in Fig.\ref{fig:Bild1} (just plotting a function) for the given conditions $I_{op}$, $B$, $h$ and $n$ and find, for a range of helium temperatures, the stable and the unstable solution $T_{cond}^{S}$ and $T_{cond}^{U}$. The energy margin is then simply the integral between $T_{cond}^{S}$ and $T_{cond}^{U}$  of some equivalent (renormalized) conductor enthalpy  $\tilde h_{cond}$ including beside conductor also some fraction of helium around the strand which has the same temperature and the same transient evolution as the strand \cite{mitchell}.  

\begin{equation}
e(I_{op},B,T_{He})=\int^{T_{cond}^S}_{T_{cond}^U}\tilde h_{cond}(T)dT
\end{equation}

For constant enthalpy, the integral in Eq.6 is proportional to $T_{cond}^U-T_{cond}^S$ and the energy margin is practically the length of the line $SU$.

How does the helium inventory in the cable influence the above picture?
It was not explicitly declared, but it is obvious from the above discussion that we considered so far that the helium temperature is not affected by the transient i.e. we assumed tacitly the case of bath cooling with helium as an infinite reservoir and not a cable-in-conduit conductor. One way to include the helium in the above analysis (as we already have done in Eq.6)is to consider that in a thin shell of helium surrounding the strand we have the same temperature as the strand while in the rest of it, referred as bulk helium, the helium temperature is considered to be constant. This is the way selected in \cite{mitchell}. The net effect is a renormalization of the conductor enthalpy ($\tilde h_{cond}$ instead of $h_{cond}$) by an amount representing the enthalpy of helium in the thin shell around the strands. Unfortunately, this correction does not solve the problem. After all, we want to describe the stability of a cable-in-conduit conductor with a finite helium reservoir where also the bulk helium temperature is changing during the pulse. The alternative to include all helium in the renormalization of the strand enthalpy would be also wrong because the temperature of the helium away from the strands is not identical with the strand temperature. The helium transients cannot therefore be neglected for a CIC conductor and we need to find the right way to include it in the analysis.

\subsection{Stability with helium transients. CICC case, 0D model}
The right way to include the helium in the stability analysis seems to be to add a second equation to Eq.4 describing the time evolution of the temperature of that part of helium, which is not included in the renormalization of $h_{cond}$ . In this case we have to solve the following 0D system of equations

\begin{eqnarray}\nonumber
(f\rho_{He}S_{He}C_{He}+\rho_{cond}S_{cond}C_{cond})\frac{\partial T_{cond}}{\partial t}=\\ \nonumber
=G-H+P_{pulse}(t)\\
(1-f)\rho_{He}S_{He}C_{He}\frac{\partial T_{He}}{\partial t}=H
\end{eqnarray}

where $f$ is the fraction of helium included in the renormalization of the conductor enthalpy and $P_{pulse}(t)$ the external heat pulse. The time dependence of the heat pulse is given by

\begin{equation}
P_{pulse}(t)=\threepartdef
{0}	{t<0}
{P}	{0<t<t_{pulse}}
{0}	{t>t_{pulse}}
\end{equation}

i.e. a simple square pulse of length $t_{pulse}$ and power $P$[W/m] or a sinusoidal $P\sin(2\pi t/t_{pulse})$ pulse. With this system of equations we first look for the existence of a steady-state solution in the absence of any perturbation. It can be shown that Eq.7, without the time dependent term (perturbation) $P_{pulse}(t)$, has no steady-state solution and that the reason is the assumed power-law for the heat generation in the strands. Indeed, a power-law conductor is permanently in current-sharing and although a current-sharing temperature $T_{cs}$ can still be defined, the heat generation is not zero below $T_{cs}$. If we look for a steady-state solution in Eq.7 with $P_{pulse}(t)=0$ then $\partial T_{cond}/\partial t$ should be zero which implies $G=H$. This has a stable solution, Eq.6, with $T_{cond}>T_{He}$ and by consequence $H=hp_{pw}(T_{cond}-T_{He})>0$. From the helium equation in Eq.7 it results that $\partial T_{He}/\partial t >0$ i.e. there is no steady-state. Reversely, if we assume that $\partial T_{He}/\partial t=0$, this implies $H=0$ i.e. $T_{cond}=T_{He}$. Then, the first line in Eq.7, gives $(f\rho_{He}S_{He}C_{He}+\rho_{cond}S_{cond}C_{cond})\partial T_{cond}/ \partial t=G>0$ and again there is no steady-state solution. This problem does not exists with the old fashioned definition of heat generation (see Appendix).

This result raises now the legitimate question if the discussion in the first section of this paper, concerning the stability is not perhaps wrong. The answer is not. First, the simple analysis with the 0D model is still correct if an infinite helium reservoir is assumed i.e. for bath cooled conductors ($T_{He}$=const). In the case of cable-in-conduit conductors with their finite (limited) helium reservoir the same analysis can be performed with some correction in the meaning and interpretation of what is meant by helium temperature in Eq.3 and 4. This is discussed below. It is clear that what is missing in Eq.7 is the fact that a) cable-in-conduit conductors are actively cooled i.e. there is a continuous advection of fresh cold helium somewhere before the section under investigation and b) that the hot helium leaves continuously the control volume, flowing in an adjacent region of the cable where either less heat generation takes place or the situation is equivalent i.e. we have same physics but different parameters like e.g. a different inlet temperature of advecting helium, a different magnetic field, etc. To keep the results of the 0D model it is therefore necessary to have in mind the following interpretations: 1) helium temperature is a parameter not a function. It is not locked at some fixed temperature e.g. at 4.2~K like in the bath cooling case, but allowed to float i.e. adapt locally to the conductor temperature and 2) the 0D model can be applied to the situation in an infinitesimal control volume placed at the most endangered section of the cable e.g. highest field or end of the high field region for a long piece of conductor. With these two conditions we can keep the interpretation of Eq.5 as a stability phase diagram also for cable-in-conduit conductors. The question is now, how to treat the transients in this case. One possible way is shown in the next section.

\subsection{Helium transients, CICC case, 1D model}
In order to take into account that a CICC is actively cooled we must extend our model to 1D. We consider a piece of conductor of length $L$ (heater length or length scale of the perturbation) exposed to a constant magnetic field $B$. Typically it is a cable section in the highest field region of a magnet. The conductor and helium temperatures are unknown functions of time and the problem is now described by the following system of equations

\begin{widetext}
\begin{eqnarray}\nonumber
(f\rho_{He}S_{He}C_{He}+\rho_{cond}S_{cond}C_{cond})\frac{\partial T_{cond}}{\partial t}
&=&\frac{\partial}{\partial x}(kS_{cond}\frac{\partial T_{cond}}{\partial x})+G-H+P_{pulse}(t) \\
(1-f)\rho_{He}S_{He}C_{He}\frac{\partial T_{He}}{\partial t}+\dot mC_{He}\frac{\partial T_{He}}{\partial x}&=&H
\end{eqnarray}
\end{widetext}

\begin{figure}[tb]
	\centering
		\includegraphics[width=6.3cm]{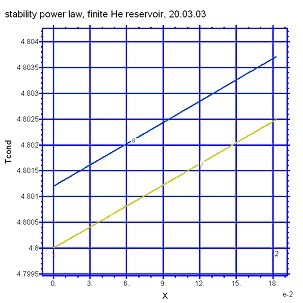}
	\caption{Conductor (blue) and helium (yellow) temperatures at steady-state before the heat pulse. Numerical solution of Eq.9 for n=15, $I_{op}$=230~A, B=6~T}
	\label{fig:Bild2}
\end{figure}

\begin{figure}[tb]
	\centering
		\includegraphics[width=6.3cm]{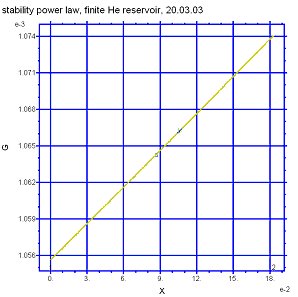}
	\caption{Heat generation and cooling. At steady-state the cooling compensates heat generation on each point along the sample length, i.e. $H(x)=G(x)$ for all $x\in [0,L]$.}
	\label{fig:Bild3}
\end{figure}

Compared to Eq.7, an advection term has been added with $\dot m$  the helium mass flow. This term describes the fact that the conductor is actively cooled. The heat conduction term in Eq.9 i.e. the longitudinal heat conduction is mandatory. It will not be considered here and is not essential for the present analysis. The effect of longitudinal thermal conduction on the steady-state was analyzed in \cite{anghel2} and as was show there it can be included in the present analysis introducing an effective (enhanced) heat exchange coefficient. The heat pulse is applied over the whole length $L$ of the sample. To this system of equation we add a Dirichlet boundary condition on the left side of the sample $T_{He}(x=0,t)=T_0$ with $T_0$ as the helium temperature at the inlet. Otherwise, we assume von Neumann boundary conditions everywhere else. The initial condition is $T_{cond}(x,t=0)=T_{He}(x,t=0)=T_0$ i.e. same conductor and helium temperatures over the sample length $L$ at $t=0$. $T_0$ will be treated as a parameter. Let us look first on what happens in the absence of the heat pulse (or equivalently at $t<0$). Because the heat generation is nonzero, the initial condition is not stable and the system evolves toward a stable steady-state solution, the solution of Eq.1 given by Eq.5, a point on the stable (left) branch of the phase diagram in Fig.\ref{fig:Bild1}. The result of a numerical calculation for a typical situation is shown in Fig.\ref{fig:Bild2}. Here we can see that at steady-state a temperature gradient develops over the length $L$ with $T_{He}(x=0,t)=T_0$ for all $t<0$ as expected, but with a conductor temperature $T_{cond}(x=0,t)>T_{He}(x=0,t)=T_0$ for all $t<0$ as given by Eq.5 at $x=0$. For $0<x<L$ we have similarly $T_{cond}(x,t)<T_{He}(x,t)$ for all $t<0$ with $T_{cond}$ given by the solutions of the equation $G(x)=H(x)$ for all $0<x<L$ as illustrated in Fig.\ref{fig:Bild3}.

What we get finally is a band of solutions describing stable steady-state conditions as pairs ($T_{cond}(x),T_{He}(x)$). This band of values is shown in Fig.\ref{fig:Bild1} by the bottom-left thick line on the stable branch of the phase diagram. 

It is obvious that the point, which is expected to quench first, is the point at $x=L$ which has the highest conductor (and helium) temperature. This will be the representative point of the system and we will concentrate from now on only on this point for the stability analysis. The position of the starting point on the phase diagram, before the pulse is applied, can be found in principle by solving the coupled algebraic-differential system of equations

\begin{eqnarray}\nonumber
G(L)&=&H(L) \\
\dot mC_{He}\frac{\partial T_{He}}{\partial x}\Bigg |_{x=L}&=&H(L)
\end{eqnarray}

which is the steady state version of Eq.9 at $x=L$. It happens that this is not necessary. Indeed, each point we can choose on the left branch of the phase diagram is a valid (and stable) initial condition as long as we look at it as a possible initial condition for the end section $x=L$ of the heated cable. Therefore, excepting the case of long-range perturbations (very large $L$) or operating current close to the DC quench condition, where the danger exist that the the cable quenches already in the steady-state, there is not necessary to solve Eq.10. Simply selecting a point on the left line in the phase diagram and looking what happens to it under a perturbation is enough. Of course, with different points we will have different helium temperatures i.e. different initial local operating temperature but this is exactly what we want and was meant by assuming floating helium temperature i.e. to study the stability under different and arbitrarily selected initial operating conditions. 

Once the issue of the initial point is decided, we turn back to the heat pulse case and try to solve the time-dependent problem expressed by Eq.9. To our knowledge, there is no other way around, the nonlinear character of this equations avoiding an analytical solution. In order to illustrate the behavior of the helium transient we solved Eq.9 numerically, neglecting the conduction term for NbTi5, one of the CONDOPT NbTi conductors \cite{wesche,bruzzone2} tested in the SULTAN facility, with the parameters shown in Table 1. The total helium mass-flow is $\dot m_{cab}$=N$_s\dot m$=5~g/s. The length of the cable is L=l$_{p}$=0.183~m and correspond to one twist-pitch of the last stage of the cable.

\begin{table}
\caption{\label{table1} Physical parameters of the CONDOPT NbTi cable used in the numerical calculation.}
\begin{ruledtabular}
\begin{tabular}{lc}
Strand type &	C\\
Strand diameter, $d_{strand}$ [mm] &	0.7\\
Cu:nCu & 7.5\\
Cable pattern & (1Cu+6)$\times3\times4\times4$\\
Number of superconducting strands, $N_s$ & 288\\
Cable diameter, $D_{cab}$ [mm] & 16.5\\
Twist pitch (last stage), $l_p$ [mm] & 183\\
Power-law index, n [-] & 15\\
non-Copper area, $S_{nCu}$ [mm$^2$] &0.2\\
Copper area, $S_{Cu}$, [mm$^2$] & 1.5\\
Helium cross-section area/strand, $S_{He}$ [mm$^2$] &0.2\\
Operating current, $I_{op}$ [A] & 230\\
External field, B [T] & 6\\
Copper density, $\rho_{Cu}$ [kg/m$^3$] & 8900\\
Helium density(4.5K, 6bar), $\rho_{He}$, [kg/m$^3$] & 130\\
Copper specific heat, $C_{Cu}$ [J/kgK] & 0.15\\
Helium specific heat, $C_{He}$ [J/kgK] & 4500\\
Total mass flow in cable, $\dot m_{cab}$ [g/s] & 5\\
Voltage criterium, $E_c$ [$\mu$V/cm] & 0.1
\end{tabular}
\end{ruledtabular}
\end{table}

The cable contains, beside the 288 superconducting strands, also 3$\times$4$\times$4=48 segregated copper strands as can be deduced from the cable pattern in Table 1. These strands are not included in the present analysis because we do not go beyond the quench point. The total helium mass flow is divided by the number of superconducting strands and gives the mass flow per strand, $\dot m$ appearing in Eq.9 and 10.
The critical current is calculated with the scaling proposed in \cite{bottura1} for NbTi. According to this scaling the current density is given by

\begin{eqnarray}\nonumber
J_c(B,T)=J_{c0}\frac{C_0}{B}\left(\frac{B}{B_{c2}}\right)^{\alpha}\left(1-\frac{B}{B_{c2}}\right)^{\beta}\times\\ \times\left[1-\left(\frac{T}{T_{c0}}\right)^{1.7}\right]^{\gamma}
\end{eqnarray}

\begin{equation}
B_{c2}(T)=B_{c20}\left[1-\left(\frac{T}{T_{c0}}\right)^{1.7}\right]
\end{equation}

where C$_0$=23.8~T, B$_{c20}$=14.5~T, T$_{c0}$=9.36~K, $\alpha$=0.57, $\beta$=0.9 and $\gamma$=1.9.
The initial condition was arbitrarily chosen to be T$_0$=4.8~K for all the results presented in this paper. This is not a special temperature. In a real magnet this would correspond to taking the region under investigation some meters away from the physical helium inlet at say 4.5K. The increase in the helium temperature from 4.5~K to 4.8~ could be the result of the index heating over this distance. Choosing another temperature, say 5.2~K would mean that we investigate a different region or the same region at a lower mass flow. 

\begin{figure}[tb]
	\centering
		\includegraphics[width=8.3cm]{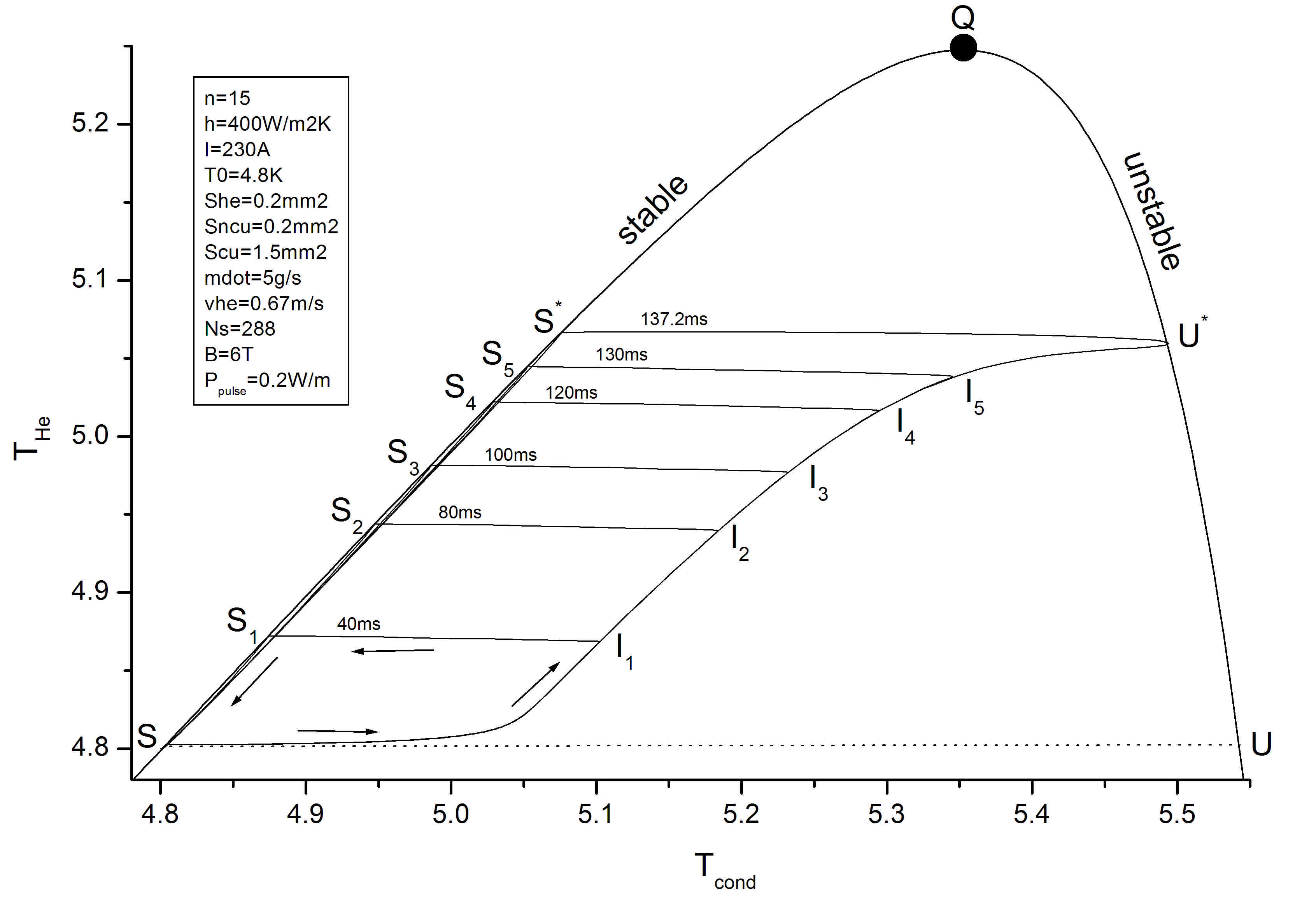}
	\caption{Transient trajectories for heat pulses with increasing duration at fixed power. The 137.2ms pulse is the last recovery pulse.}
	\label{fig:Bild4}
\end{figure}

\begin{figure}[tb]
	\centering
		\includegraphics[width=8.3cm]{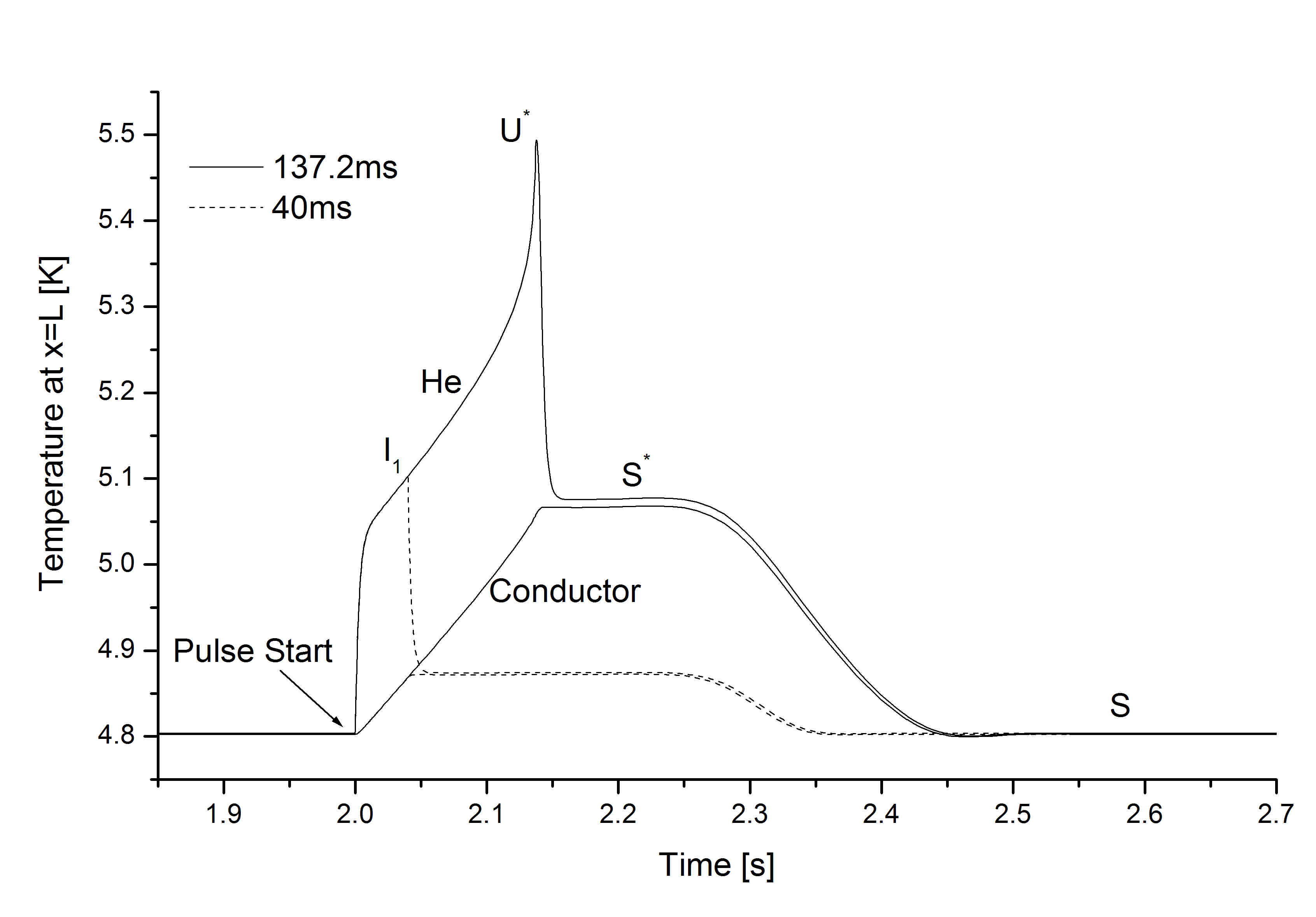}
	\caption{Time dependence of the conductor and helium temperatures for a recovery pulse (40ms) and the last recover pulse (137.2ms)}
	\label{fig:Bild5}
\end{figure}

In a first run we applied at t=2~s a pulse of 0.2~W/m with increasing duration beginning with 40~ms. Eliminating the time between the solutions $T_{cond}(x=L,t)$ and $T_{He}(x=L,t)$ we get a manifold of points which can be plotted in the phase diagram of Fig.\ref{fig:Bild1} as a line. We call this line the trajectory of the transient. Some results are presented in Fig.\ref{fig:Bild4} and Fig.\ref{fig:Bild5}. The general behavior is like this: starting at t=0 with equal helium and conductor temperatures, T$_0$=4.8~K, the system evolves first toward a stable steady-state S with the conductor temperature a little bit higher than that of the helium. This would be the point we could choose from the beginning if we had solved Eq.1 before the heat pulse. When the heat pulse is activated we observe first a rapid increase of the conductor temperature followed later by an increase in the helium temperature. At the end of the pulse, point I$_1$, the conductor temperature starts to decrease rapidly, accompanied again by a slightly helium temperature increase until the stable line in the phase diagram is reached at a point S$_1$ placed above S.

Finally, a simultaneous, equilibrium cooling of helium and conductor takes place with the representative point moving down along the stable phase line, back to point S. This is due to the advection of cold helium in the region under investigation (the effect of the thermal perturbation is wash-out). The stability issue is decided by the position of the trajectory point at the end of the pulse i.e. if this point did or did not touch the unstable branch of the phase diagram. As can be seen in Fig.\ref{fig:Bild4} for pulses with duration up to 130ms, the last point of the pulse: I$_1$, I$_2$, I$_3$, etc. is still away from the unstable line but it is coming closer and closer to it until finally a last small increase of the pulse duration from 130~ms to 137.2~ms is enough to bring the trajectory almost in touch with what we can call now "the quench line". This is still a recovery but already at 137.3~ms there is indeed a quench as shown  by the sudden breaking of the convergence in the numerical calculation (crash of program). The first lesson we can learn from this analysis is that the interpretation of Eq.5 (or Fig.\ref{fig:Bild1}) as a stability phase diagram is not lost when we take into account the helium transients. The phase diagram does not loose its validity. We have a left branch of stable steady-state points, which can be called the "recovery line", and a right branch that can be called the "quench line", both separated by the dc quench point Q. The only difference is that when applying a perturbation the trajectory does not touch the quench line at the point U, the projection of the initial point S on the unstable branch of the phase diagram, as in the case discussed in Fig.\ref{fig:Bild1}(complete neglect of Helium) but somewhere higher at the point U$^*$ due to the over-heating of helium and as a manifestation of helium finite inventory. How much higher, depends on the pulse duration, power and pulse form as we will show later. Because the slope of the quench line is large and grows with increasing n, we will have $T_{cond}^U \cong T_{cond}^{U^*}$ in the limit of very large n. The energy of the pulse goes practically into the Helium whose temperature increases from T$_{He}^S$ to T$_{He}^{S^*}$.

\begin{figure}
	\centering
		\includegraphics[width=8.3cm]{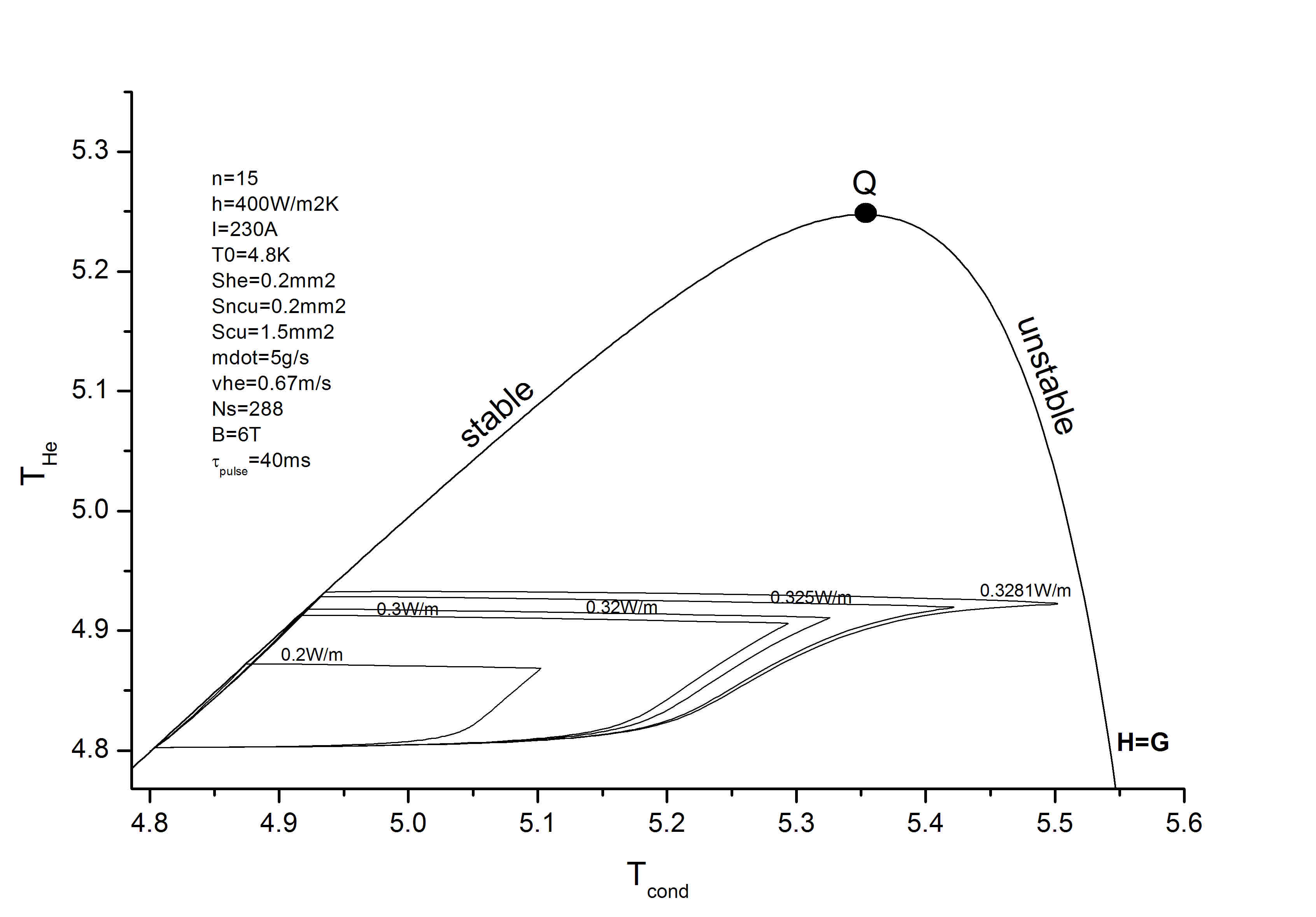}
	\caption{Transients trajectories for pulses with increasing power and fixed duration. The lasr recovery pulse has a power of 0.3281~W/m. At 0.3282~W/m the conductor quenches.}
	\label{fig:Bild6}
\end{figure}

Let us look now at the case of a pulse of fixed duration and increasing power. This a case is illustrated in Fig.\ref{fig:Bild6} where the power of a 40~ms pulse was increased in 0.1~W/m steps starting at 0.2~W/m until a quench occurs. The last recovery pulse is for 0.3281W/m and the first quench occurs at 0.3282~W/m. Observe the very fine difference, making the exact positioning on the quench line rather difficult. The physics is the same as before. During the pulse, the conductor temperature starts to increase with almost stagnant helium temperature, then, with some delay, the helium temperature starts to increase too due to the heat transfer from strand to helium and the point I is reached with the maximum conductor temperature. After the pulse, the conductor temperature drops rapidly while the helium temperature continues to increase slightly (almost negligible) until the quasi-stable point S$^*$ is reached. From now on cooling by the fresh cold helium is the dominant mechanism (thermal wash-out) and the conductor and helium temperature evolve, quasi-stationary, back to the initial point S following closely the stable branch of the phase diagram.

From both analyzes and many others not shown here, the following conclusions can be drawn: a) The phase diagram described by the line  does not loose its meaning by considering the helium transients. b) Neglecting the helium transient leads to a wrong (over) estimation of the stability margin for cable in conduit conductors. Indeed, neglecting the helium inventory would give a stability margin quantified by the length of the line SU in Fig.\ref{fig:Bild4} which spans a larger temperature range as the range spanned by the line S$^*$U$^*$ when the helium is taken onto account. The stability margin without the helium transient would be

\begin{eqnarray}\nonumber
e_{margin}=e_{cond}=(f\rho_{He}S_{He}C_{He}+\\
+\rho_{cond}S_{cond}C_{cond})(T_{cond}^U-T_{cond}^S)
\end{eqnarray}

Here we have a larger temperature difference but only a fraction $f$ of helium is taken into account. The correct value of the energy margin is given by

\begin{eqnarray}\nonumber
e_{margin}=e_{cond}+e_{He}=(f\rho_{He}S_{He}C_{He}+\\
+\rho_{cond}S_{cond}C_{cond})(T_{cond}^{U^*}-T_{cond}^S)+\\ \nonumber
(1-f)\rho_{He}S_{He}C_{He}(T_{He}^{S^*}-T_{He}^S)
\end{eqnarray}

Now the conductor temperature difference is lower but the enthalpy change of helium is added to the total energy. c) the position of point S$^*$ on the phase diagram can be estimated with the help of the following equation

\begin{equation}
T_{He}^{S^*}\cong T_{cond}^{U^*}\approx T_0+\frac{P_{pulse}t_{pulse}}{\rho_{He}S_{He}C_{He}}
\end{equation}

if the pulse power is known. For the two cases analyzed here we get with this equation, $T_{He}^{S^*}$=5.034~K (Fig.\ref{fig:Bild4}) and $T_{He}^{S^*}$=4.912~K (Fig.\ref{fig:Bild6}), values very close to the real values 5.066~K and 4.933~K (point S$^*$).
d) For the same conditions of current and field the energy margin is not unique. For pulses of fixed power and variable duration the energy margin is $e^*=(Pt_{pulse})_{margin}$=27~mJ/m  and for pulses of fixed duration and variable power it is 13~mJ/m. The energy margin depends on the pulse characteristic like power, time and form. This is one of the main results of this paper. 
  
\begin{figure}[tb]
	\centering
		\includegraphics[width=8.3cm]{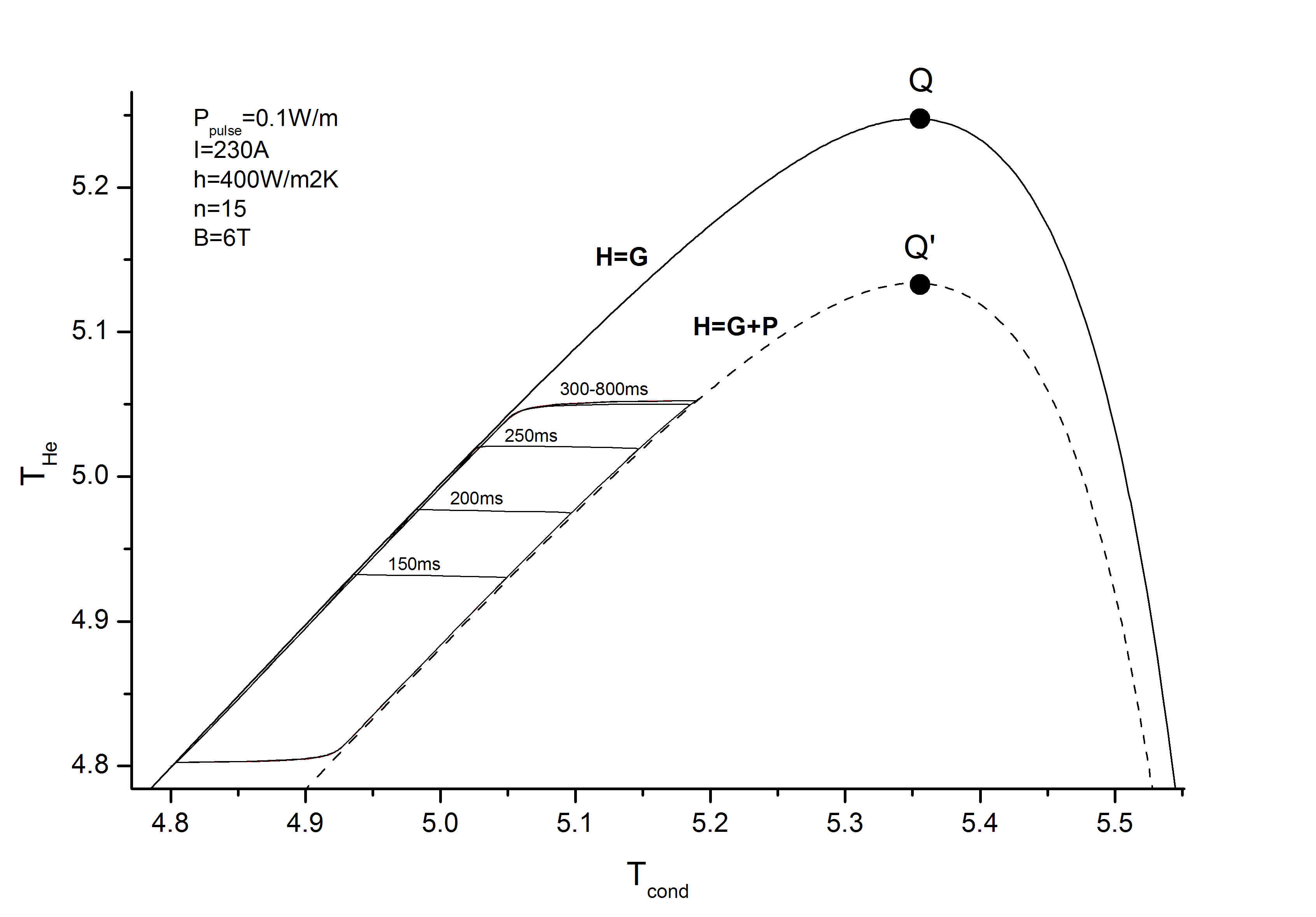}
	\caption{Phase diagram for  low energy pulses. Pulse power is 0.1W/m}
	\label{fig:Bild7}
\end{figure}

An interesting situation arises if one considers low energy pulses with variable pulse duration. One generally expects that with increasing pulse duration a quench will be reached sooner or later. Surprisingly, this is not the case as shown in Fig.\ref{fig:Bild7}. Here we have the situation with a pulse power of 0.1~W/m and the trajectories for increasing pulse duration, starting at 150~ms are shown. As can bee seen, after an initial increase of the final conductor temperature proportional to the pulse duration as expected, a saturation effect appears at about 300~ms. For all pulse durations larger that 300~ms, all the trajectories are overlapping. 

\begin{figure}
	\centering
		\includegraphics[width=8.3cm]{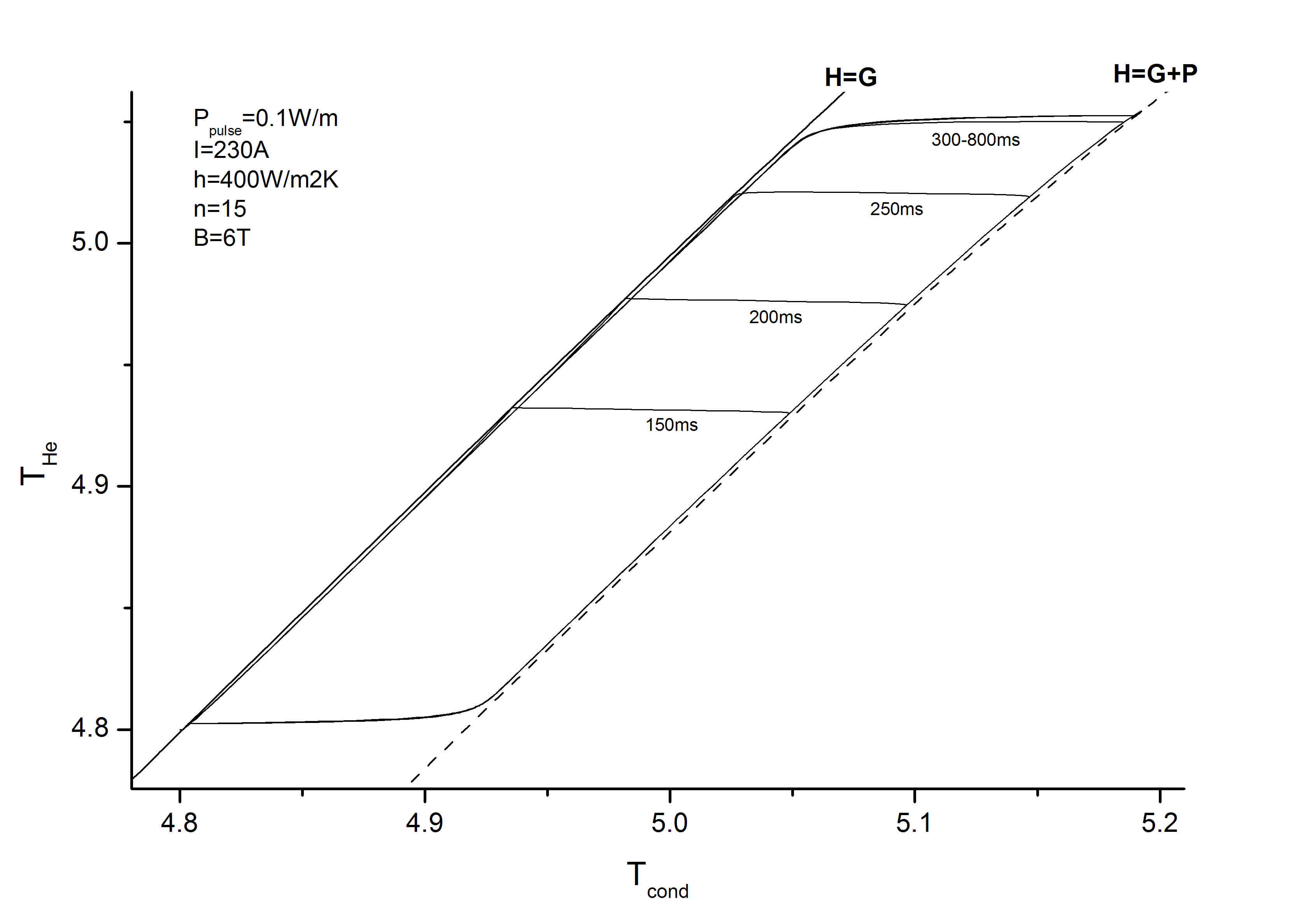}
	\caption{A close-up view on the left-side of the phase diagram in Fig.\ref{fig:Bild7}, showing pulses with non-contacting trajectory for pulse duration below 300~ms.}
	\label{fig:Bild8}
\end{figure}

The explanation of this unexpected phenomenon is as follows. For not too high pulse energies (we will discuss later the limit) the system can be in steady state not only in the absence of the pulse but also with the pulse as even if it would be applied forever ($t_{pulse}=\infty)$. Therefore we must consider beside the phase line given by $G$=$H$ also the phase line given by $G+P$=$H$ as indicated by the dashed-line in Fig.\ref{fig:Bild7}. Now it can be seen that for pulse duration up to 300~ms, the trajectory never touches the new phase line although it comes very close to it. But beyond 300~ms, the trajectory touches the new phase line on some point on the "stable" side as viewed better in Fig.\ref{fig:Bild8} which shows a close-up in the region of interest. During the pulse, at long durations, a new steady-state condition is reached corresponding now to the solution of the new equation $G+P=H$. After the pulse, the system will return to the stable-state given again by the old equation $G=H$. For this energy, and for lower energies of course, the system will never quench no matter how long the pulse duration is.

\begin{figure}[tb]
	\centering
		\includegraphics[width=8.3cm]{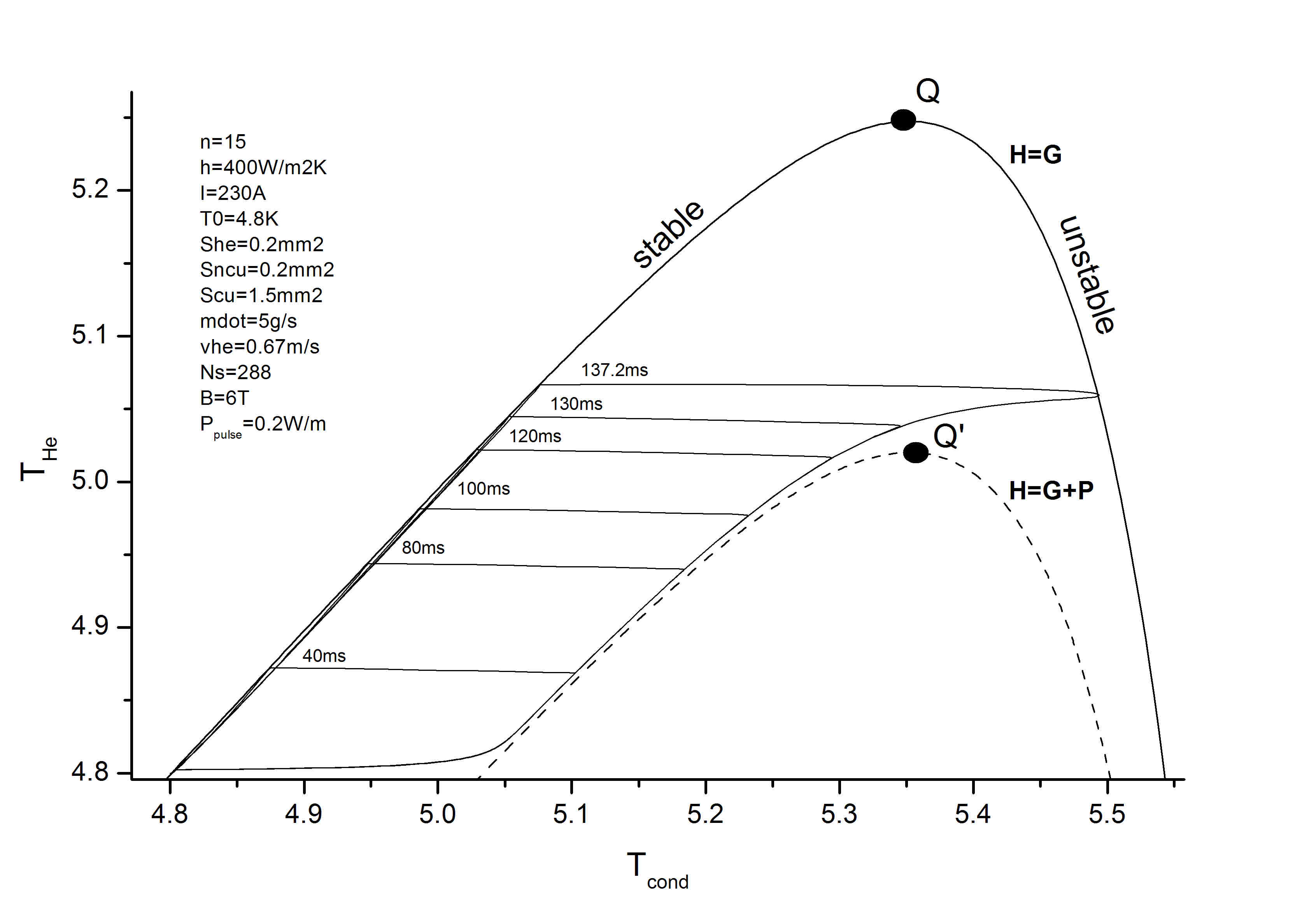}
	\caption{Replica of Fig.\ref{fig:Bild4} showing that for a pulse of 0.2~W/m there is a quench before the trajectory has the oportunity to touch the second stability line $H=G+P$}
	\label{fig:Bild9}
\end{figure}

It happens that there is a threshold for this effect. As shown in Fig.\ref{fig:Bild9}, if the pulse power is a little bit higher e.g. for 0.2~W/m the transient is such that there is no contact between the trajectory and the new phase line including the pulse power. A quench is possible in this case. I do not have found yet an analytical solution but the exact mathematical formulation for the condition of this threshold is given by the pulse energy such that the first contact point between the trajectory of the transient and the new phase line $G+P=H$  is exactly at the apex point Q'. Indeed, if the two lines did not cross already before the apex of $G+P=H$ there is no chance to do it later since the slope of the new phase line became negative beyond the apex while the slope of the trajectory is always positive.  

We have mentioned before, that for low energy pulses, the conductor can be in a steady state not only without the pulse power but also with the pulse power applied forever. It seems that a second threshold exists here. With increasing $P$, the new phase line is moving down, until the apex point reaches the line $T_{He}=T_{He}^S$. Then, at pulse powers higher than this threshold it become impossible for any trajectory to ever touch the phase line $H=G+P$ for any possible pulse duration.

\begin{figure}[tb]
	\centering
		\includegraphics[width=8.3cm]{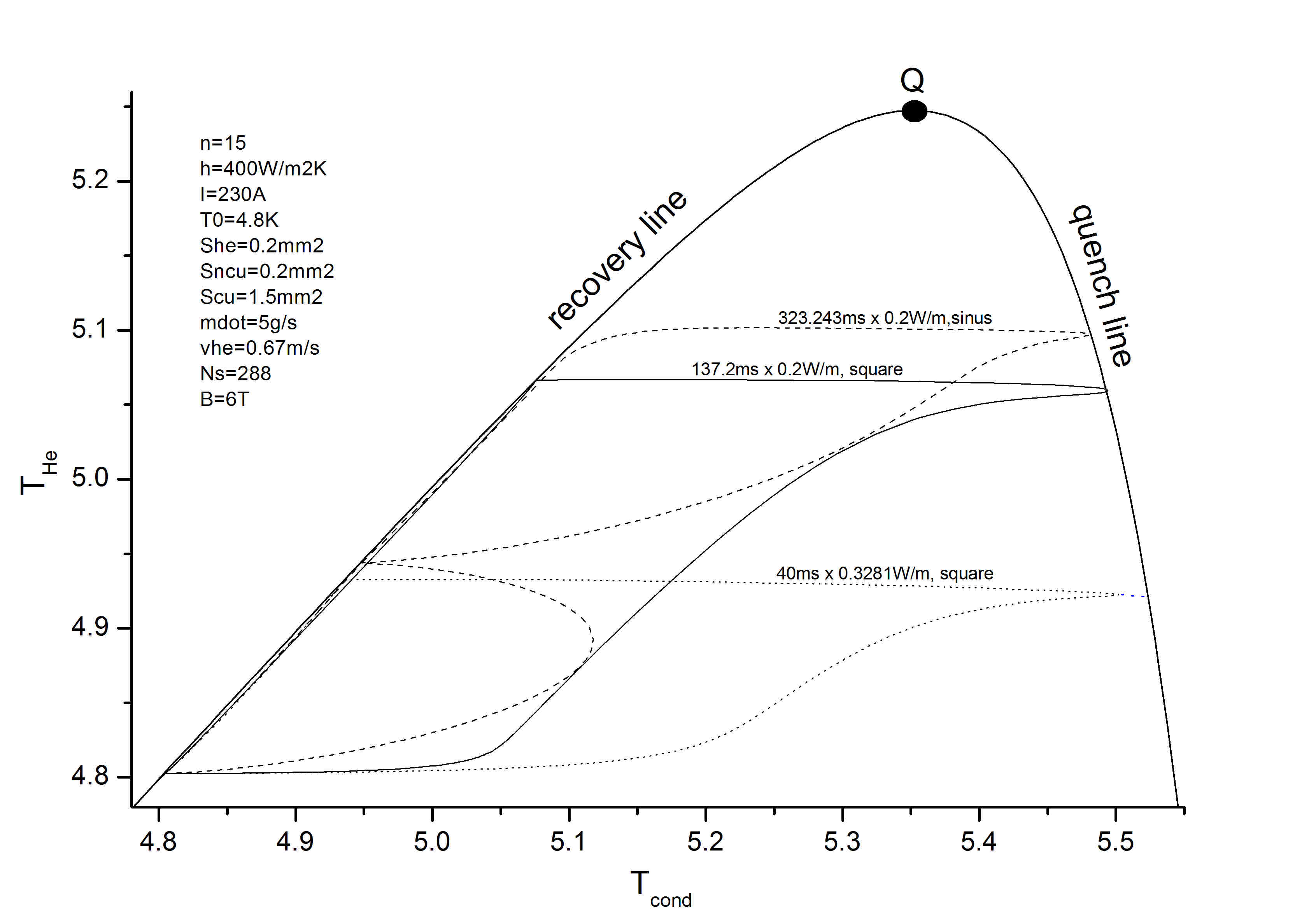}
	\caption{Trajectories of the last recovery pulse for three different pulses with different power, pulse time and form. This shows that the energy margin is not uniquely defined.}
	\label{fig:Bild10}
\end{figure}

\subsection{Conclusions and discussions}
Some features of stability of cable-in-conduit conductors with power-law heat generation were investigated. A stability phase diagram has been introduced represented by Eq.5. The phase line has a stable branch (the recovery line) to the left of the quench point and an unstable branch (the quench line) on the right side of the quench point. The apex Q of the phase line is the DC quench-point. We introduced the trajectory of a control point at the end of the heated zone and used it to investigate the transient behavior of the conductor. A conductor is stable under a given perturbation (it recovers) if the trajectory forms a closed cycle starting and ending on the stable line without cutting the quench line. The last recovery trajectory defines the stability margin of the conductor. Some general features of the stability of cable-in-conduit conductors for perturbations with constant power and variable duration and constant duration and variable power were investigated. The main and the most important result is that there is no unique energy margin for given initial conditions. The energy margin depends on the pulse characteristics like power, pulse time and pulse form as shown in Fig.\ref{fig:Bild10}. It was shown that for pulses of fixed power and increasing pulse duration a power threshold exists. For a pulse power below this threshold the conductor never quenches, even for infinite pulse duration. We have shown that this is linked to the fact that the transient trajectory touches a second stability line given by $G+P=H$. This seems to happen always when the line S$^*$U$^*$ cuts the second phase line. 

From the phase diagram it can be seen that the energy margin for a given pulse is a continuous decreasing function either of helium temperature (the operating temperature) at fixed current or operating current at fixed helium temperature. There is no limiting current in this model as opposed to the conclusions in the previous stability models \cite{ciazinski,bottura2}. 
The mass flow dependence of stability enters not only through the heat exchange coefficient but also and more intricately trough the change in the initial steady-state condition and cooling conditions during the pulse. The steady-state conductor and helium temperature at the end of the pulse are dependent on the mass flow. 

This study was mainly devoted to the development of general concepts and methods. We investigated only the most general properties of the stability of cable-in-conduit conductors with power-law current-voltage characteristic. Details such as self-field effects (peak field), dependence on the power-law index (including the dependence on critical current), the effect of superconductor to copper and inter-strand current redistribution, imposed inhomogeneous current distribution, the helium mass flow effect, the transient effects on the heat exchange coefficient and the length-scale of the perturbation can be investigated in the same framework. Excepting the inter-strand current redistribution and the imposed inhomogeneous current distribution, their treatment is trivial and does not bring any substantial change in the concept. 

The peak-field effect, for example, can be considered simply by drawing the phase line for $B=B_{peak}=B_0+kI_{op}$ and the sc/copper current sharing by using the parallel resistor approximation leading to a slightly modified formula for the index heating function $G=G_{sc}(I_{sc})+G_{Cu}(I_{Cu})$  with the condition $I_{sc}+I_{Cu}=I_{op}$ . The results presented here are valid only for the range of currents and/or temperatures where the power-law description of real conductors makes sense i.e. where an index $n$ can be defined and measured. 

Finally, we mention that the transient behavior close to the dc quench point can be also analyzed using the method developed in \cite{rachmanov} for the particular case of a bath-cooled composite HTS conductor.

\appendix
\section{Appendix}
It is interesting to note and compare the power-law heat generation with the conventional (traditional, old) way of describing the heat generation in a superconductor, based on the current-sharing temperature concept. In this case the heat generation function is defined by

\begin{widetext} 
\begin{equation}
G(T_{cond},I_{op})=\threepartdef
{0}	{T_{cond}<T_{cs}}
{\dfrac{\rho_{Cu}I_{op}^2}{S_{Cu}}\left[\dfrac{T_{cond}-T_{cs}}{T_c-T_{cs}}\right]}	{T_{cs}<T_{cond}<T_c}
{\dfrac{\rho_{Cu}I_{op}^2}{S_{Cu}}}	{T_{cond}>T_c}
\end{equation}
\end{widetext}

with $\rho_{Cu}$ and $S_{Cu}$ the copper resistivity and cross-section area, $T_c$ -the critical temperature and $T_{cs}$ -the current-sharing temperature. A steady-state solution of Eq.7 allways exists for $T_{cond}=T_{He}<T_{cs}$ since at $T_{cond}<T_{cs}$, $G=0$ from Eq.9 and $H=0$ if $T_{cond}=T_{He}$. This is practically the definition of current-sharing temperature $T_{cs}$ i.e. the temperature above which the heat generation starts to be non-zero.

\section{Acknowledgments}

\begin{acknowledgments}
We would like to express our acknowledgment G. Pasztor for valuable discussions and support.
\end{acknowledgments}


\end{document}